\documentclass[a4paper]{article}
\usepackage{pdfsync}
\usepackage[utf8]{inputenc}
\usepackage{graphicx}
\usepackage{amssymb,amsfonts,amsmath,amsthm}
\usepackage{hyperref}
\usepackage{graphicx}
\usepackage{multirow}
\usepackage{verbatim}
\usepackage{cite}
\usepackage{color}

\theoremstyle{plain}

\theoremstyle{definition}

\newtheorem{rmk}{Remark}[section]

\begin{document}

\title{\textbf{A probabilistic model for crystal growth applied to protein
  deposition at the microscale}}


\author{V. J. Bol\'os$^1$, R. Ben\'{\i}tez$^1$, A. Eleta-L\'opez$^2$, J. L. Toca-Herrera$^3$ \\ \\
{\small $^1$ Dpto. Matem\'aticas para la Econom\'{\i}a y la Empresa, Facultad de Econom\'{\i}a,} \\
{\small Universidad de Valencia. Avda. Tarongers s/n, 46022 Valencia, Spain.} \\
{\small e-mail\textup{: \texttt{vicente.bolos@uv.es}\qquad \texttt{rabesua@uv.es}}} \\ \\
{\small $^2$ Self-Assembly Group, CIC nanoGUNE, } \\
{\small Tolosa Hiribidea 76, 20018, Donostia/San Sebasti\'an, Spain.} \\
{\small e-mail\textup{: \texttt{a.eleta@nanogune.eu}}} \\ \\
{\small $^3$ Department of NanoBiotechnology, Institute for Biophysics,}\\
  {\small University of Natural Resources and Applied Life Sciences (BOKU).}\\
  {\small Muthgasse 11, 1190 Vienna, Austria.} \\
{\small e-mail\textup{: \texttt{jose.toca-herrera@boku.ac.at}}} \\}

\date{February 2018}

\maketitle

\begin{abstract}
A probabilistic discrete model for 2D protein crystal growth is
  presented. This model takes into account the available space and can
  describe growing processes of different nature due to the
  versatility of its parameters which gives the model great
  flexibility. The accuracy of the simulation is tested against a real
  protein (SbpA) crystallization experiment showing high agreement
  between the proposed model and the actual images of the nucleation
  process. Finally, it is also discussed how the
  regularity of the interface (i.e. the curve that separates the crystal
  from the substrate) affects to the evolution of the simulation.
\end{abstract}

\section{Introduction}
\label{intro}

A crystal is a (three dimensional) periodic
arrangement of repeating ‘structural motifs’, which can be atoms,
molecules, or ions \cite{Kittle2007}. Technically, the process in which a
(small) crystal becomes larger is called crystal
growth. Crystallization is commonly referred to the fact of creating
nucleation points first, followed by crystal growth. Crystal growth is
important for the understanding of biomineralization, molecular
diffusion and adsorption, or fractal formation \cite{Veis2013,Huang17,Semenova93}. Theoretical crystallization models distinguish between thermodynamic and kinetic
conditions \cite{Himawan2006,Zhou2008}. For problems concerning growth rate and growth from limited resources, the Avrami \cite{Avrami1,Avrami2} and the Gompertz functions can be used, respectively, for data modelling \cite{Laird1964}.

Two dimensional in-vivo protein crystallization is commonly utilized
to study the function and structure of proteins \cite{DOYE200640}. In-situ crystal
growth can also be used to test growth and molecular adsorption
models. In this work we have taken advantage of a bacterial protein,
SbpA from \textit{Lysinibacillussphaericus} CCM2177, which forms crystalline
arrays on different interfaces \cite{SLEYTR2001231}. In particular, SbpA is able to
self-assemble from solution building square lattices on silicon
dioxide and self-assembly monolayers \cite{Aitzibe-2010}. Former studies on SbpA
crystallization indicate on one hand that the process consists of a
transition from amorphous to crystalline state \cite{Chung16536}, and on another
hand, that protein-substrate determines the properties of the
crystalline domain \cite{Lejardi13}.  

In this work we propose an approach to describe and model 2D protein
crystal growth. The structure of the paper is as follows. In Section
\ref{sec:1} we present the probabilistic model and define all the parameters
involved in it. This model takes into account the available space for
growing and also reproduces different shapes and porosity (or
lacunarity). In Section \ref{sec:fit} we analyze the different
parameters and how they affect the development of the simulation.
All these different tunable parameters give the model a very high
flexibility and allow us to successfully simulate real protein
crystallization processes. Finally, in Section \ref{sec:results} we
test the model by simulating the growth of a SbpA crystal. We conclude
the paper discussing how the regularity of the interface affects the
evolution of the crystallization process.

\section{The model}
\label{sec:1}

The model we present here is a discrete-space discrete-time model for
a square region in which a regular $n\times n$ square mesh is defined. At each
step $k=1,2,\ldots $, one cell is filled by the crystal, and the corresponding time is
denoted by $t_k$, with $t_0=0$ the initial time. In our
model, $t_k$ is computed at each step $k$ and it determines only the growth rate of the crystal, but
it does not affect to the structure, i.e. its shape.

\subsection{Structure}

The space filled by the crystal when the $k$th cell is occupied
is given by the
\textit{occupation matrix} $M^{(k)}$, defined by $M^{(k)}_{ij} = 1$ if cell $(i,j)$
is occupied, and $M^{(k)}_{ij} = 0$ if it is free.

In this way, we also define a \textit{probability matrix} $P^{(k)}$ so
that the probability of the cell $(i,j)$ to be occupied at the $(k+1)$th
step is given by $P^{(k)}_{ij}$. Nevertheless, in the simulation
procedure we shall use a \textit{relative probability matrix}
$C^{(k)}$ which gives the number of chances that has a cell to be
occupied, and it is related with $P^{(k)}$ in this way:
\[
P^{(k)} = \frac{C^{(k)}}{\sum_{i,j} C^{(k)}_{ij}}.
\]
For example, if we want that each cell has exactly the same
probability of being occupied at the beginning, then we can set
$C^{(0)}_{ij}=1$. On the other hand, if we want the cells on a
particular region to be twice as likely to be occupied at the
beginning, then we can set $C^{(0)}_{ij}=2$ for such cells and
$C^{(0)}_{ij}=1$ otherwise. Moreover, we can also force to have, for
example, only one crystallization nucleus taking $C^{(0)}_{ij}=0$
except for one cell, as it is done in Figure \ref{fig:kkbc} with a
cell in the middle of the square region.

In our model, proteins will depose onto the substrate, occupying
cells, following the next three structure rules:
\begin{enumerate}
\item If a cell is occupied, it cannot be occupied again, i.e. if
  $M^{(k)}_{ij} = 1$, then $C^{(k)}_{ij} = 0$.
\item If a cell is occupied, then the probability of occupation of the
  adjacent free cells is increased, see \eqref{eq:2}.
\item The probability of occupation of a free cell depends also on the
  available area in a neighborhood of that cell, see \eqref{eq:2} and \eqref{eq:fijk}.
\end{enumerate}

This rules are independent of $t_k$ and hence they only
determine the structure of the crystal, not its growth rate.
Note that the first rule implies that this is a fully 2D model, i.e. no
height increase is considered here.

Also note that, taking into account the second rule, the space scale
is much larger than the characteristic length of the crystal
structure. That is, each cell does not represent a single crystal, but
a larger quantity. In fact, the scale is such that one occupied cell
only increases the probability of occupation of the adjacent cells.

In our model, we propose that the chance of occupation of a free cell $(i,j)$ at the $k$th step is given by
\begin{equation}
  \label{eq:2}
  C_{ij}^{(k)} = \left( C_{ij}^{(0)}+\rho \cdot \left(\mathcal{A}_{ij}^{(k)}\right)^\beta\right) \cdot \mathcal{F}_{ij}^{(k)},
\end{equation}
where $\rho $ is a positive constant, $\mathcal{A}_{ij}^{(k)}$ is the number of adjacent occupied cells, and $0<\mathcal{F}_{ij}^{(k)}\leq 1$ is a function
related to the available area surrounding the free cell (see \eqref{eq:fijk}). The real parameter $\beta $ determines the importance
of the adjacent occupied cells and it is usually set to $1$ (see Section \ref{sec:fit}).
Note that, if a free cell has not any adjacent occupied cell, then its chance of occupation is $C_{ij}^{(k)} = C_{ij}^{(0)} \cdot \mathcal{F}_{ij}^{(k)}$.

\begin{rmk} In order to speed up the algorithm, the third rule can be modified
by only considering free cells adjacent to an occupied cell and so, equation \eqref{eq:2} would only apply to free cells with $\mathcal{A}_{ij}^{(k)}>0$. Hence, the chance of occupation of a free cell with no adjacent occupied cells (i.e. $\mathcal{A}_{ij}^{(k)}=0$) would be
$C_{ij}^{(k)} = C_{ij}^{(0)}$ instead of $C_{ij}^{(0)} \cdot \mathcal{F}_{ij}^{(k)}$. It gives practically
the same results when the number of crystallization nuclei is low and the algorithm is much faster.
\end{rmk}

\begin{figure}[htb]
  \centering
  \includegraphics[width=\linewidth]{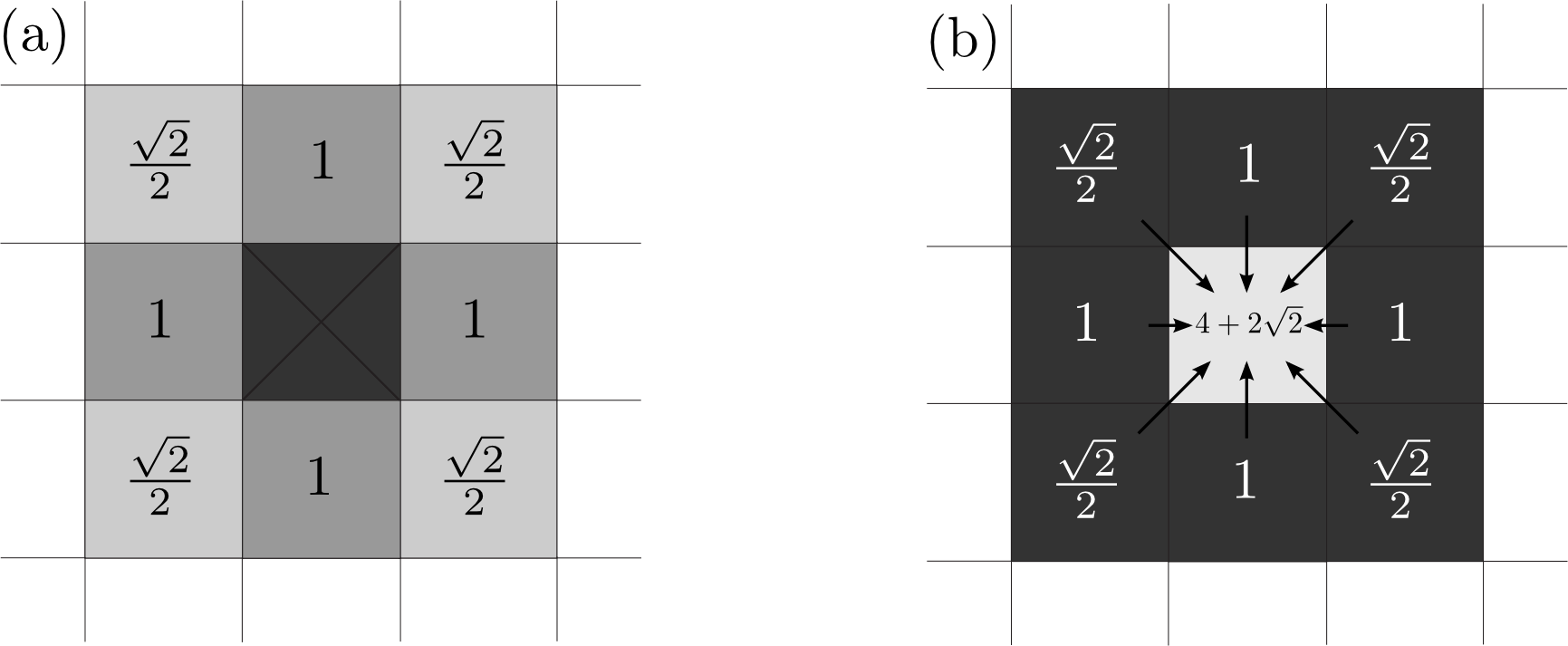}
  \caption{(a) Influence of an occupied cell (center) on the adjacent
    free cells. The weight in the diagonal directions is $\sqrt{2}/2$
    for avoiding square-shaped evolutions.  (b) For a fully surrounded
    free cell (center), its total number of occupied adjacent cells
    is $4+2\sqrt{2}$.}
  \label{fig:adjcells}
\end{figure}

For determining $\mathcal{A}_{ij}^{(k)}$ we have to take into account
that in the diagonal directions there is a weight factor of
$\sqrt{2}/2$ (see Figure \ref{fig:adjcells}). This weight factor is
set in order to avoid square-shaped evolution produced by the
discretization in space, where circular ones should appear.

\begin{figure}[htb]
  \centering
  \includegraphics[width=0.5\linewidth]{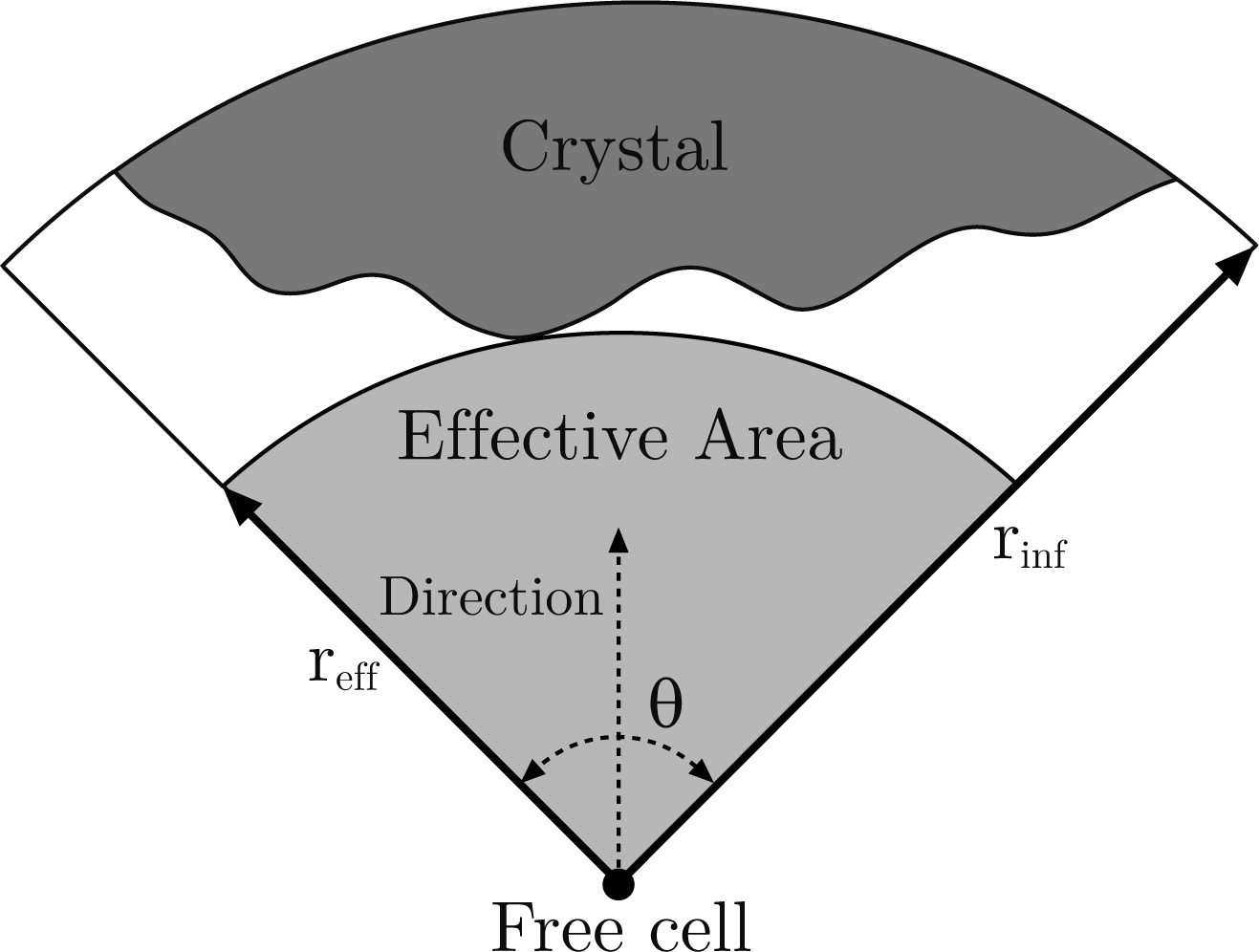}
  \caption{Effective area of a free cell in the north direction with an opening angle $\theta =\pi /2$.}
  \label{fig:eff}
\end{figure}

With respect to $\mathcal{F}_{ij}^{(k)}$, first we have to define some concepts. The \textit{effective radius}, $r_{\textrm{eff}}$, of a free cell in a given direction with an \textit{opening angle} $\theta $ is the distance from the cell to the nearest occupied cell lying in the range of this direction. Then, the corresponding \textit{effective area} is the area of the circular section with radius $r_{\textrm{eff}}$ and angle $\theta$ (see Figure \ref{fig:eff}). The underlying idea is that the effective area is a zone where the free cell can capture proteins, but in this model, it is very simplified.

Although $\theta $ could be a parameter, in our model we take $\theta =\pi /2$ because it greatly simplifies the discretized algorithm and it produces results matching the real measures (see Section \ref{sec:fit}).
Moreover, we only take into account effective radii lesser than a given parameter called \textit{influence radius}, $r_{\textrm{inf}}$, beyond which, occupied cells do not hinder the crystal growth. So, if $r_{\textrm{eff}}>r_{\textrm{inf}}$ (even $r_{\textrm{eff}}=+\infty $) then we take $r_{\textrm{eff}}=r_{\textrm{inf}}$. Next, we define the \textit{maximum effective radius} $R_{ij}^{(k)}$ of a free cell $(i,j)$ at the $k$th occupation stage as the maximum of the corresponding effective radius, but only in the eight main directions (for operational purposes): east, northeast, north, northwest, west, southwest, south and southeast.

Hence, for a free cell, we define
\begin{equation}
\label{eq:fijk}
\mathcal{F}_{ij}^{(k)} = (1-\alpha )+\alpha \left( \frac{R_{ij}^{(k)}}{r_{\textrm{inf}}}\right) ^{\gamma },
\end{equation}
where $0<\alpha \leq 1$, $\gamma >0$, and hence, $0<\mathcal{F}_{ij}^{(k)}\leq 1$. The parameter $\alpha $ determines the importance of the maximum effective radius in the model and it is usually taken close to $1$. Setting $\alpha =0$ means that the probability of occupation of a cell does not depend on the maximum effective radius and then, rule 3 is not satisfied. On the other hand, we usually set $\gamma =2$ because in this case $\left( R_{ij}^{(k)}/r_{\textrm{inf}}\right) ^2$ is the ratio between the effective area and the maximum possible effective area (i.e. the area of the circular section of radius $r_{\textrm{inf}}$ and angle $\theta $). For convenience, $\mathcal{F}_{ij}^{(k)}=0$ for an occupied cell. With this statement, rule 1 is a consequence of equation \eqref{eq:2}.

\subsection{Kinetics}\label{sec:kinetics}

Let $A(t)$ be the area defined by the crystal measured in occupied cells. Note that, although in this model $A(t)$ is a step function, in this work we model the physical phenomena in which $A(t)$ can be considered as a derivable real variable function. Nevertheless, we are going to consider the occupied area proportion
\begin{equation}
\label{eq:y}
y(t)=A(t)/n^2,
\end{equation}
that ranges from $0$ to $1$. In this section, we are going to set $A(0)=0$ and hence $y(0)=0$. Moreover, $A'(0)$ will denote the number of expected occupied cells per time unit at $t=0$. It is not too important for the kinetics in a qualitative sense because, for example, a simulation with $A'(0)=20$ will produce practically the same results as with $A'(0)=10$ (leaving all other parameters unchanged), but in half the time.

With respect to the crystal growth rate, we can fit the the Avrami function, designed for modeling crystallization and some chemical reactions \cite{Avrami1,Avrami2}:
\begin{equation}
\label{eq:avrami}
y(t)=1-e^{-\kappa t^{\eta}},
\end{equation}
where $\kappa >0$ and $\eta \in \left\{ 1,2,3,4\right\} $. However, in our case, where the crystal growth is 2-dimensional, $\eta $ can be considered equal to $1$.
Regarding $\kappa $, it is given by the initial crystal growth rate, since
\begin{equation}
\label{eq:kappa}
\kappa =A'(0)/n^2.
\end{equation}

The Avrami function fits well in the first stages of the crystal growth, but when the free space resource becomes scarce there are bigger differences (see Figure \ref{fig:imagesInforme2}) since the Avrami function does not take into account this kind of resource.
To solve this problem in the last stages, we can fit the Gompertz function, designed for growth with limited resources (in our case, the free space) \cite{Laird1964}:
\begin{equation}
\label{eq:gompertz}
y(t)=e^{-b e^{-ct}},
\end{equation}
where $b,c>0$. But note that this function can only model the crystal growth rate in an advanced state, since the initial occupied area in the Gompertz model is not zero.
So, it can be done a piecewise fit using the Avrami function \eqref{eq:avrami} for the first stages, and the Gompertz function \eqref{eq:gompertz} for the rest, when the lack of free space becomes significant.

Finally, as an alternative for the piecewise fit, we can use a unique ``hybrid'' Avrami-Gompertz function:
\begin{equation}
\label{eq:avrami-gompertz}
y(t)=1-e^{-\kappa t e^{-ct}},
\end{equation}
where $\kappa ,c>0$. Note that \eqref{eq:avrami-gompertz} can only be applied in a time interval $\left[ 0,T\right] $ with $T<1/c$ because, according to \eqref{eq:avrami-gompertz}, $y'(t)<0$ for $t>1/c$. Moreover, the parameter $\kappa $ in \eqref{eq:avrami-gompertz} is also equal to $A'(0)/n^2$, as in \eqref{eq:avrami}, but the parameter $c$ in \eqref{eq:avrami-gompertz} is not the same as in \eqref{eq:gompertz}.

\section{The parameters}
\label{sec:fit}

There are two kinds of parameters: \textit{structural} and \textit{kinetic}. The structural parameters determine the shape of the crystal, while the kinetic parameters are responsible for the crystal growth rate. We should also mention the parameter $n$ (size of the square mesh) which is of great importance since its value affects both, the structure and the kinetics of the crystal growth.

\subsection{Structural parameters}

\begin{itemize}

\item $\rho $: the ``occupation chance multiplier'' for free cells adjacent to an occupied one, see \eqref{eq:2}. Mainly, it is responsible for the number of crystallization nuclei along the process (see Figure \ref{fig:nuclei}).

\begin{figure}[htb]
  \centering
  \includegraphics[width = 0.7\linewidth]{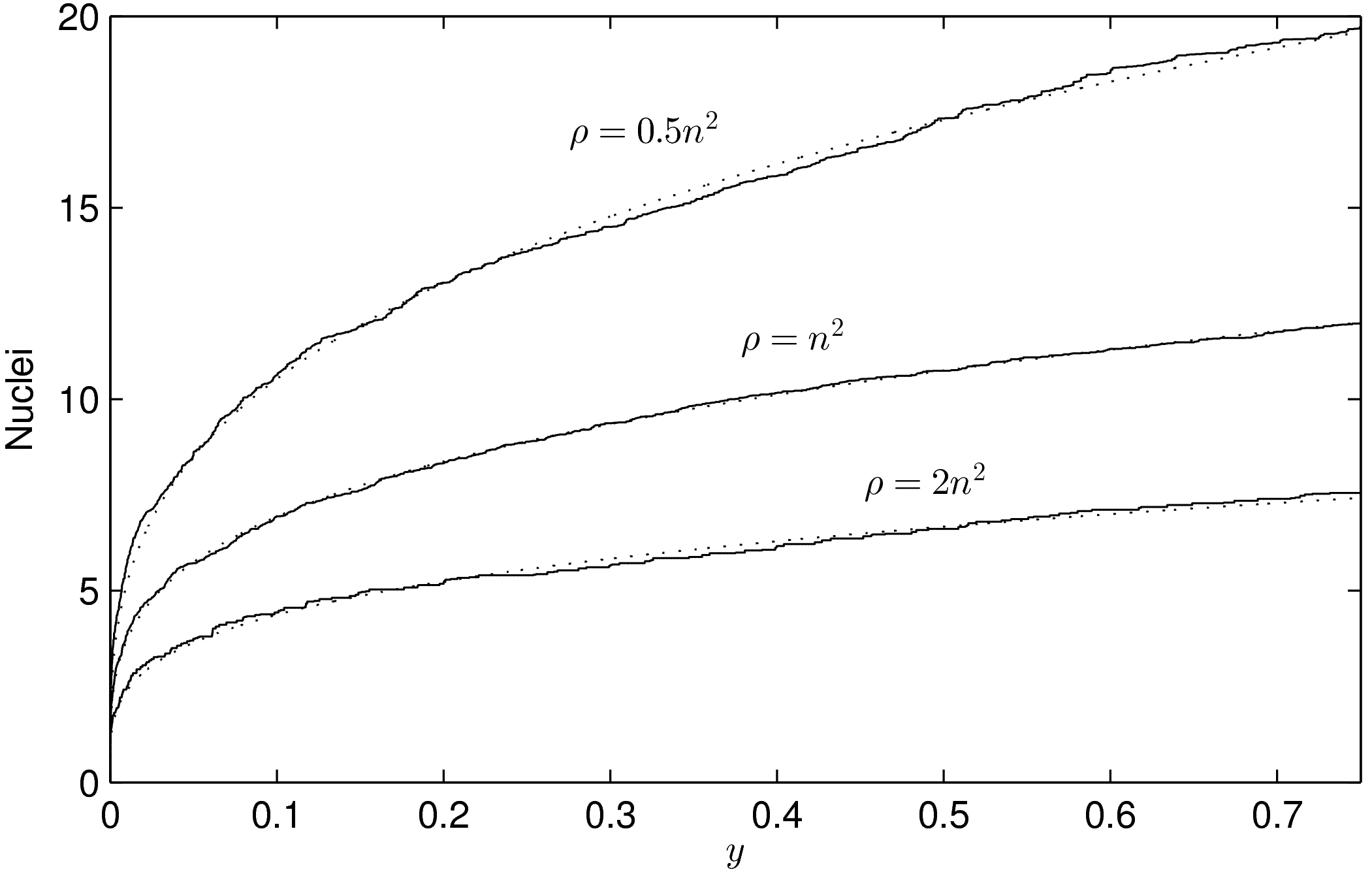}
  \caption{Number of crystallization nuclei along the process, from the beginning ($y=0$) to $75\%$ of total occupied area ($y=0.75$), for different values of $\rho$. We represent  (solid lines) the means of $100$ simulations for $\rho =0.5n^2$, $\rho = n^2$ and $\rho =2n^2$ respectively, with $n=256$ and $C^{(0)}_{ij}=1$ for all cell $(i,j)$. The rest of the parameters are $\beta =1$, $\alpha = 0.8$ and $r_{\textrm{inf}}=32$, but they are not determinant. It is shown (dotted lines) that the number of crystallization nuclei follows a power law of the form $ay^b$ with $a=21.41$, $b=0.3083$ ($\rho =0.5n^2$), $a=12.98$, $b=0.2722$ ($\rho =n^2$), and $a=8.006$, $b=0.2628$ ($\rho =2n^2$). Note that parameter $a$ represents the expected number of nuclei at the end of the process ($y=1$) and, from this result, it seems to hold $a_1\approx 1.635 a_2$ for $\rho_1=\frac{1}{2}\rho_2$.}
  \label{fig:nuclei}
\end{figure}

\item $\theta $: the opening angle, see Figure \ref{fig:eff}. In our model $\theta =\pi /2$ and so, it is not taken as a parameter. Nevertheless, it can be changed causing effects on the crystal shape: the greater the angle is, the bigger the ``fjords'' are (see Figure \ref{fig:theta180}).

\begin{figure}[htb]
  \centering
  \includegraphics[width = 0.6\linewidth]{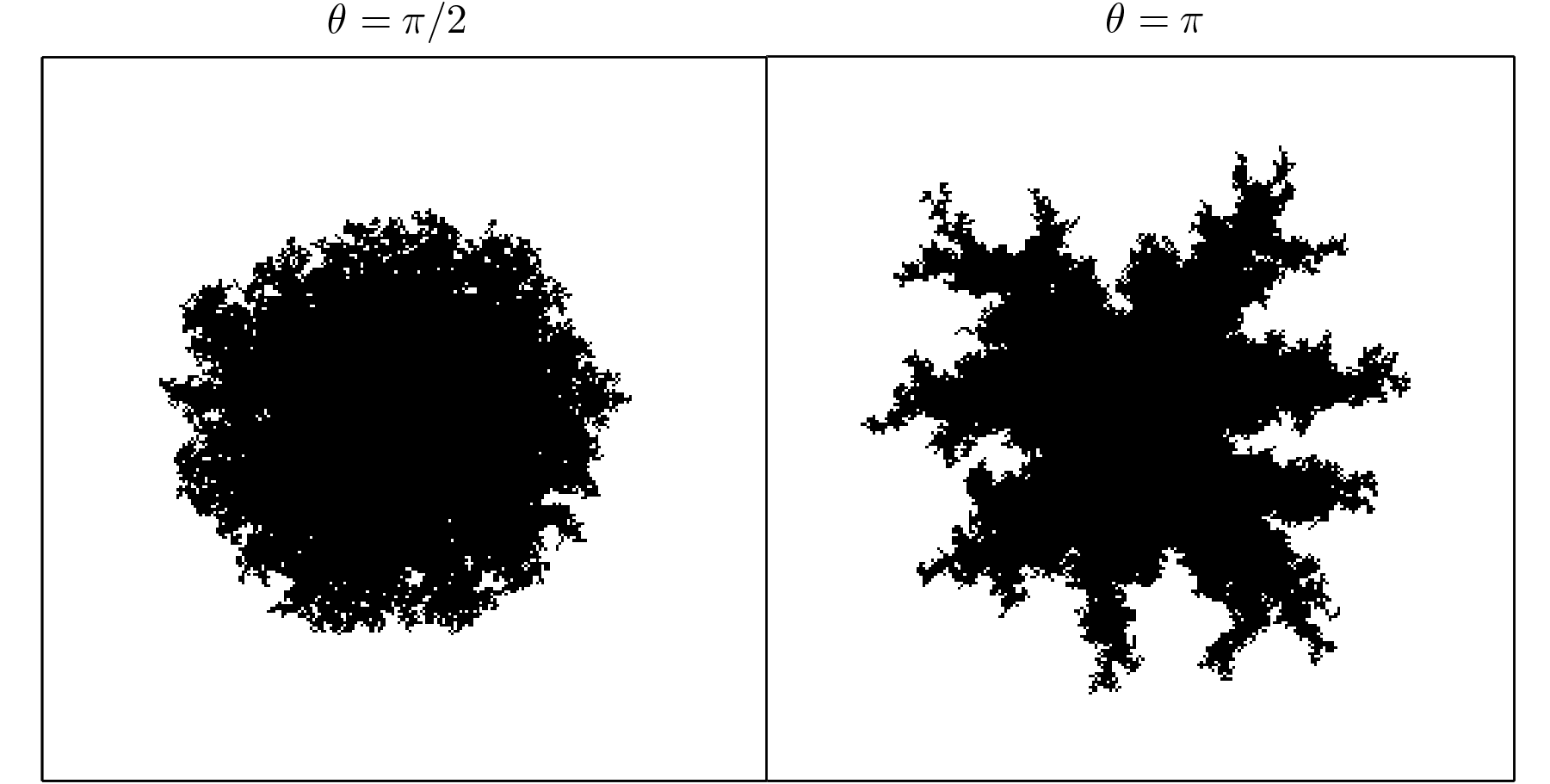}
  \caption{Shape at $y=0.25$ (i.e. $25\%$ of total occupied area, see
    \eqref{eq:y}) of $1$ crystallization nucleus for opening angle
    $\theta =\pi /2$ (left) and $\theta =\pi $ (right), taking
    $\alpha = 0.8$, $\beta = 1$, $r_{\textrm{inf}}=32$ and
    $n=256$. The greater the angle $\theta $ is, the bigger the
    ``fjords'' are.}
  \label{fig:theta180}
\end{figure}

\item $\beta $: the ``sponginess'' parameter, see
  \eqref{eq:2}. Determines the effect of adjacent occupied cells on a
  free cell, and it is usually set to $1$. In Figure \ref{fig:kkbc} it
  is shown how this parameter (along with $\alpha $) affects the shape
  of the crystal.

\begin{figure}[htb]
  \centering
  \includegraphics[width = \linewidth]{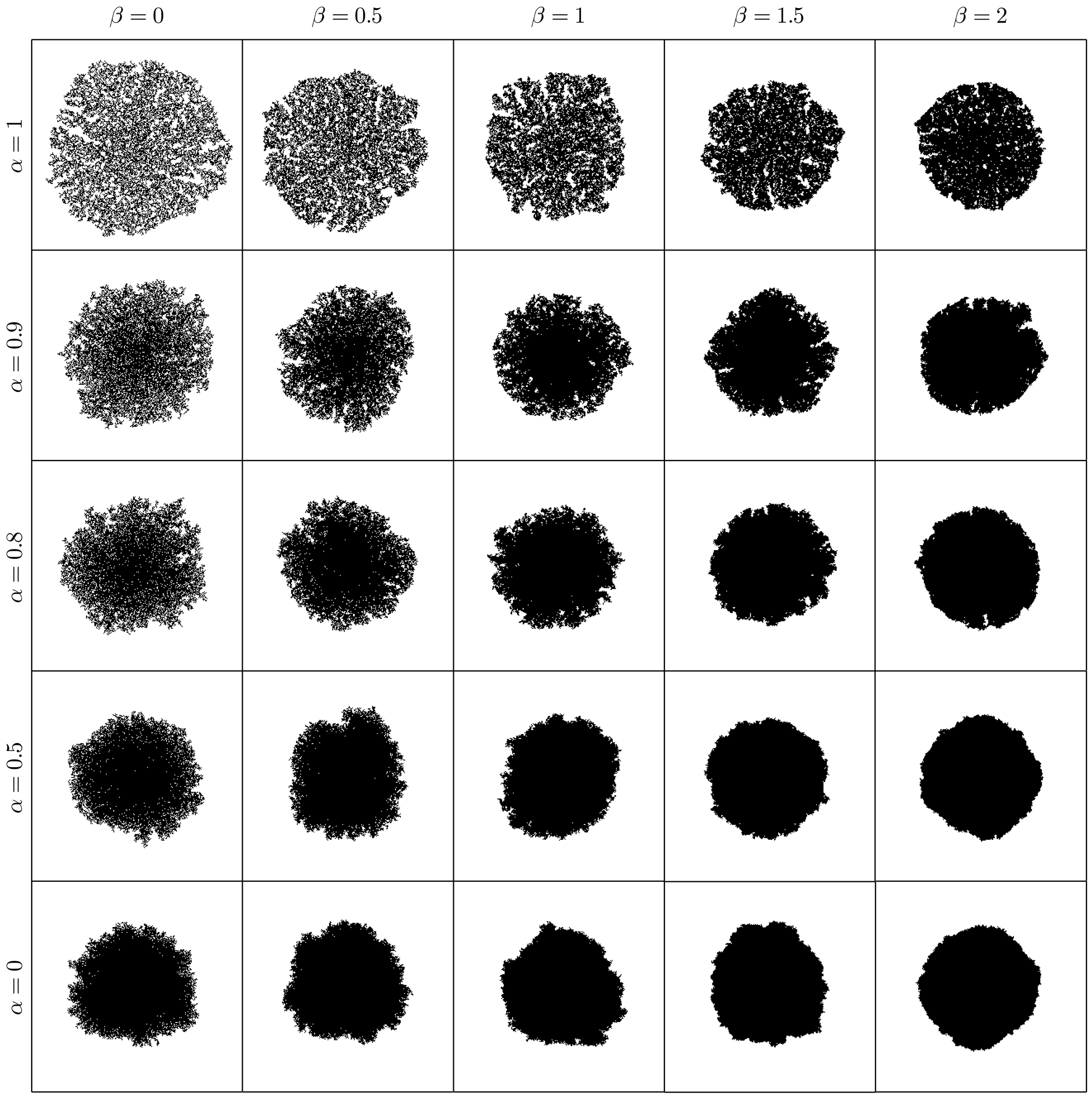}
  \caption{Shape at $y=0.25$ (i.e. $25\%$ of total occupied area, see
    \eqref{eq:y}) of $1$ crystallization nucleus for different values
    of $\alpha $ and $\beta $, taking $n=256$. The parameter $\beta $
    determines the ``sponginess'' of the crystal, and $\alpha $
    represents the ``difficulty of filling''. We have taken
    $r_{\textrm{inf}}=32$, although it is not determinant in these
    figures.}
  \label{fig:kkbc}
\end{figure}

\item $\alpha $: the ``difficulty of filling'' parameter, see
  \eqref{eq:fijk}. It ranges from $0$ to $1$ and determines the
  importance of the maximum effective radius in the model. It is also
  related to the rate at which void regions are filled. Setting
  $\alpha =0$ means that the probability of occupation of a cell does
  not depend on the maximum effective radius. In Figure \ref{fig:kkbc}
  it is shown how this parameter (along with $\beta $) affects the
  shape of the crystal.

\item $r_{\textrm{inf}}$: the influence radius, see Figure
  \ref{fig:eff}. Determines the range at which occupied cells hinder
  the crystal growth, and so the width of the ``channels'' (see Figure
  \ref{fig:file3focos}).

\begin{figure}[htb]
  \centering
  \includegraphics[width = \linewidth]{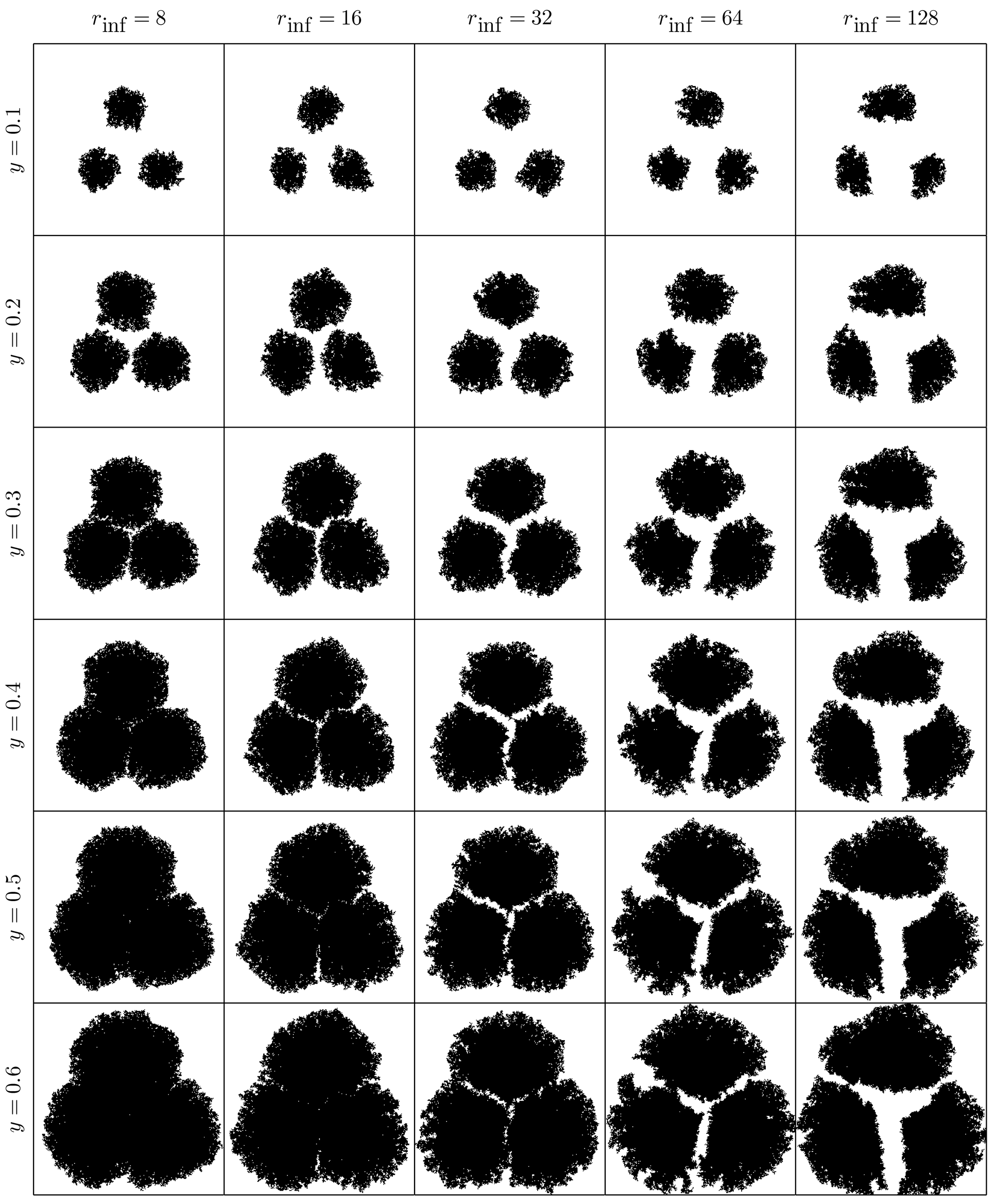}
  \caption{Evolution of $3$ crystallization nuclei (for occupied area
    proportion from $y=0.1$ to $y=0.6$) with different values of
    $r_{\textrm{inf}}$, taking $\alpha = 0.8$, $\beta = 1$ and
    $n=256$. The parameter $r_{\textrm{inf}}$ determines the width of
    the ``channels''.}
  \label{fig:file3focos}
\end{figure}

\item $\gamma $: the ``effective dimension'' parameter, see
  \eqref{eq:fijk}. It must be positive and, as it is explained before,
  in our model it is set to $2$. Also determines how the maximum
  effective radius affects the crystal growth.

\end{itemize}

\subsection{Kinetic parameters}

\begin{itemize}

\item $\kappa, \eta$: Parameters of the Avrami function
  \eqref{eq:avrami}, that models the first stages of the process. In
  our case, where the crystal growth is 2-dimensional, it can be
  considered $\eta =1$. On the other hand, $\kappa $ is determined by
  the initial crystal growth rate $A'(0)$ (see \eqref{eq:kappa}).

\item $b,c$: Parameters of the Gompertz function \eqref{eq:gompertz},
  that models the last stages of the process.

\end{itemize}

\section{Results}
\label{sec:results}
\subsection{Protein recrystallization}
\begin{figure}[htb]
  \centering
  \includegraphics[width = \linewidth]{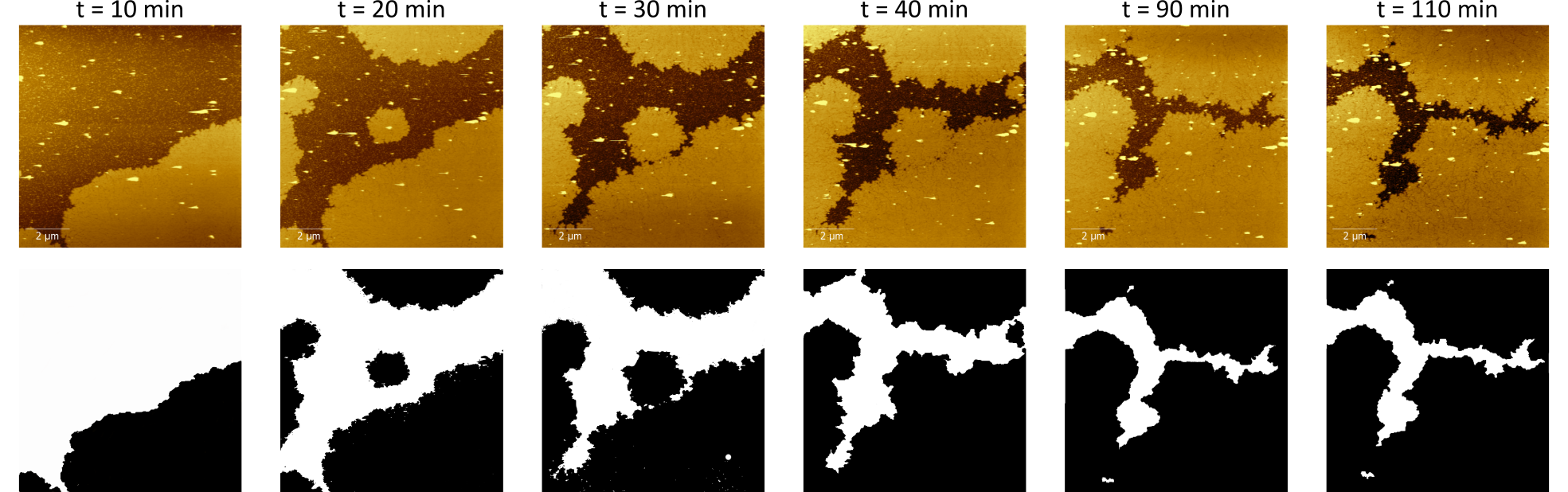}
  \caption{SbpA crystal growth sequence at different times. After the last time stamp at $t=110$ minutes, the crystal was
    stabilized, growing very slowly until the complete filling of the
    substrate more than 12 hours after the beginning.}
  \label{fig:imagesInforme}
\end{figure}

In order to validate the model we tested it against the growth of a
bacterial which crystallizes forming structures called S-layers
(SbpA). This SbpA protein was recrystallized on a SiO$_2$ substrate
and the overall process was scanned by an AFM obtaining height
images (see \cite{Aitzibe-2010} for the specific details of the
experiment). In Figure \ref{fig:imagesInforme}, the first row of images
shows the crystallization process for times ranging from 10 to 110
minutes. The crystal growth continues steadily until more than the 98\%
of the substrate is covered 12 hours later.

Taking the $10\mu m\times 10\mu m$ images of the crystal deposition at
different time stamps, the total surface occupied by the crystal was
estimated by transforming the images to black and white 8-bits images
and using a self-developed MATLAB code which computes the number of black
pixels over the total amount of pixels of the image. The second row
of images in Figure \ref{fig:imagesInforme} depicts, as an
example, the transformation into black and white images of the
original images shown in the top row.

\begin{figure}[htb]
  \centering
  \includegraphics[width = 0.8\linewidth]{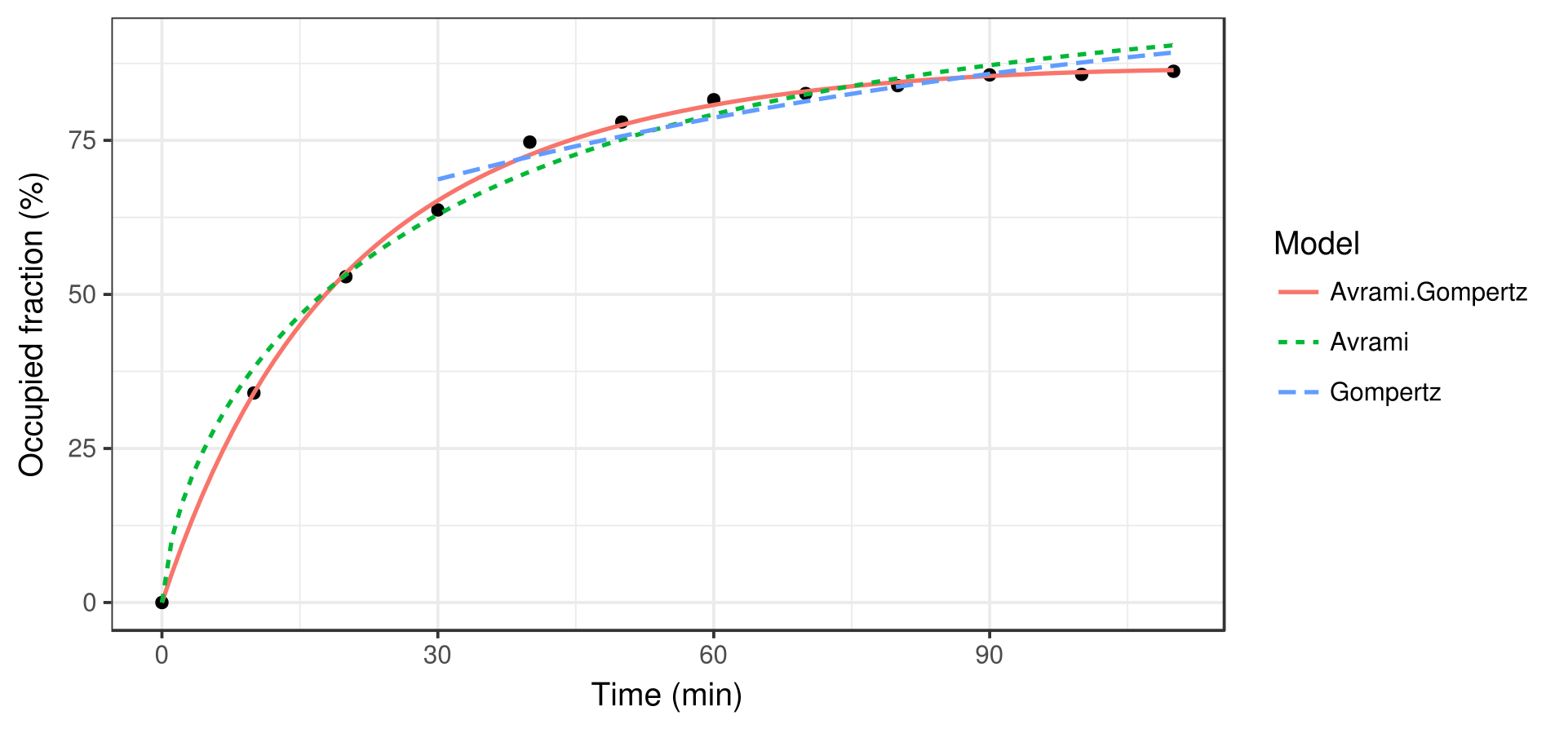}
  \caption{Estimation of the occupied fraction at different time stamps, from $t = 0$ to $t = 120$ minutes (black dots) together with the three different kinetic growth models: Avrami, Gompertz and Avrami-Gompertz.}
  \label{fig:imagesInforme2}
\end{figure}

Once the images are transformed and the occupied fraction is estimated, a growth curve is fit to the data
according to the growth models described in Section
\ref{sec:kinetics} (see Figure \ref{fig:imagesInforme2}). It is noteworthy how the Avrami model fits the
data accurately at the beginning, but the growth rate for later
moments is too high because it does not consider the limitations of
space. As the crystal grows, the available space is reduced and so is
the crystal growth rate. On the other hand, the Gompertz model
captures the slow growth rate from the first 30 minutes, but it is not
suitable for the first stages of the crystallization process. In fact,
since the Gompertz growth model does not vanish at $t=0$, it cannot be
applied there. Finally, the Avrami-Gompertz model is able to
accurately fit the growth rate of the recrystallization process both
at the beginning and at the final steady state moments. The nonlinear
least square fits were performed using the R statistical software
\cite{Rsoft} and the Levenberg-Marquardt algorithm provided by the
\texttt{minpack.LM} package \cite{minpack}.

\begin{figure}[htb]
  \centering
  \includegraphics[width = 0.95\linewidth]{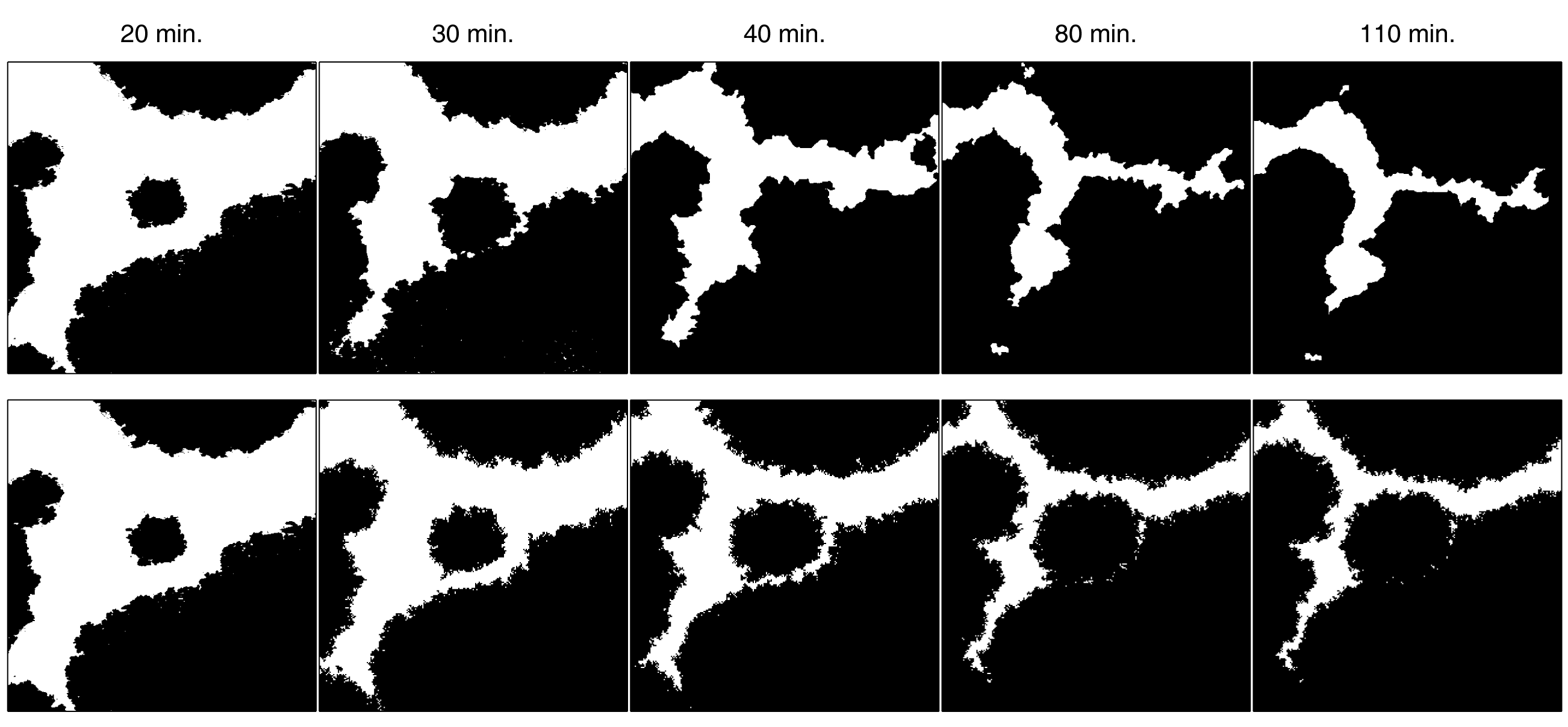}
  \caption{Comparison between real images (up) and simulation (down)
    at different times, with initial data corresponding to the real
    data at 20 minutes. The values of the structural parameters taken
    in the simulation are $\beta = 1$, $\alpha = 0.5$ and
    $r_{\textrm{inf}}=n/8$ (with $n=256$). The kinetic parameters have
    been set to fit the real data, according to the Avrami-Gompertz
    model \eqref{eq:avrami-gompertz}.}
  \label{fig:imagesAitziber}
\end{figure}

In order to assess the validity of the model as a description of the
recrystallization process at the micro scale, we run several
simulations using as initial condition the black and white image of the
crystallization at $t = 20$ minutes. The choice of this initial
condition is based on the fact that, with a 50\% of occupied fraction,
despite the stochastic nature of the model, the simulation would give
results that could be compared to the experimental data. Figure
\ref{fig:imagesAitziber} shows the evolution of the recrystallization
process for one of the runs of the simulation (all of them gave
results which were indistinguishable at a glance). The top row shows
the black and white pictures obtained from the original AFM images while the second row
depicts the simulation results at the same occupation fractions. The
equivalence between the occupation fraction and the time stamp was
obtained by the Avrami-Gompertz kinetic model. The initial (leftmost)
pictures are the same in both rows. It is remarkable the close agreement
between the simulation results and the experimental data. 

\subsection{Influence of the contour regularity}

Once the growth model has proven suitable for describing the protein
recrystallization process, the question of the dependence of the
growth rate on the regularity of the interface contour arises. The
rationale behind all this is based on the fact that every system will
evolve so its free energy is minimized. In crystal growth processes,
the energy is highly dependent on the surface tension, or the line
tension (for a 2D interface) \cite{barabasi_fractal_1995}.  Therefore,
the higher the length of this tension line is, the higher the free
energy will be. Thus if the system seeks to minimize the free energy
in the fastest  possible way, the crystal should grow the most in the
parts were the interface line is longer. Since the length of a curve
is closely related to its regularity, we should find that a crystal
with an irregular contour grows faster (under the same conditions)
than a crystal with a regular one.

\begin{figure}[htb]
  \centering
  \includegraphics[width = \linewidth]{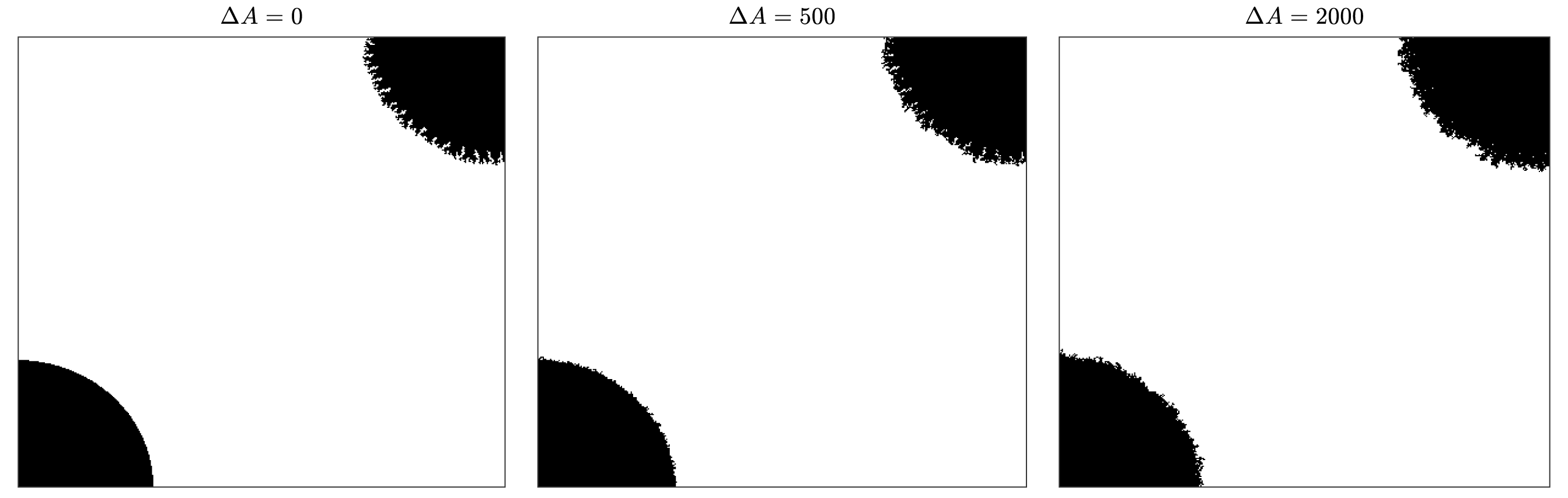}
  \caption{Evolution of two nuclei (irregular and regular), initially
    both with $10000$ occupied cells. The values of the structural
    parameters taken in the simulation are $\beta = 1$, $\alpha = 0.5$
    and $r_{\textrm{inf}}=n/8$ (with $n=400$). We have taken
    $C^{(0)}_{ij}=0$ for cells with no adjacent occupied cells to
    prevent the formation of new nuclei.}
  \label{fig:fractal_c05}
\end{figure}

In order to assess how the regularity of the contour affects the growth
process in our model, a simulation experiment was conducted with the
following initial condition: a regular and an irregular nucleus both
with the same size (measuring 10000 occupied cells) were placed in
opposite corners of a square big enough so the existence of any of the
two nuclei did not affect the evolution of the other (see Figure
\ref{fig:fractal_c05} left picture). We run 2000 iterations of the
simulation setting the parameters to $\beta = 1$, $\alpha = 0.5$ and
$r_{\text{inf}} = 50$. We also set $C^{(0)}_{ij} = 0$ for cells with
no adjacent occupied cells so no new nuclei could appear. Figure
\ref{fig:fractal_c05} shows how the regular nucleus (bottom left
corner) grows slower than the irregular nucleus (top right corner) and
how its contour becomes less and less regular as the crystal
deposition continues.

\begin{figure}[htb]
  \centering
  \includegraphics[width = \linewidth]{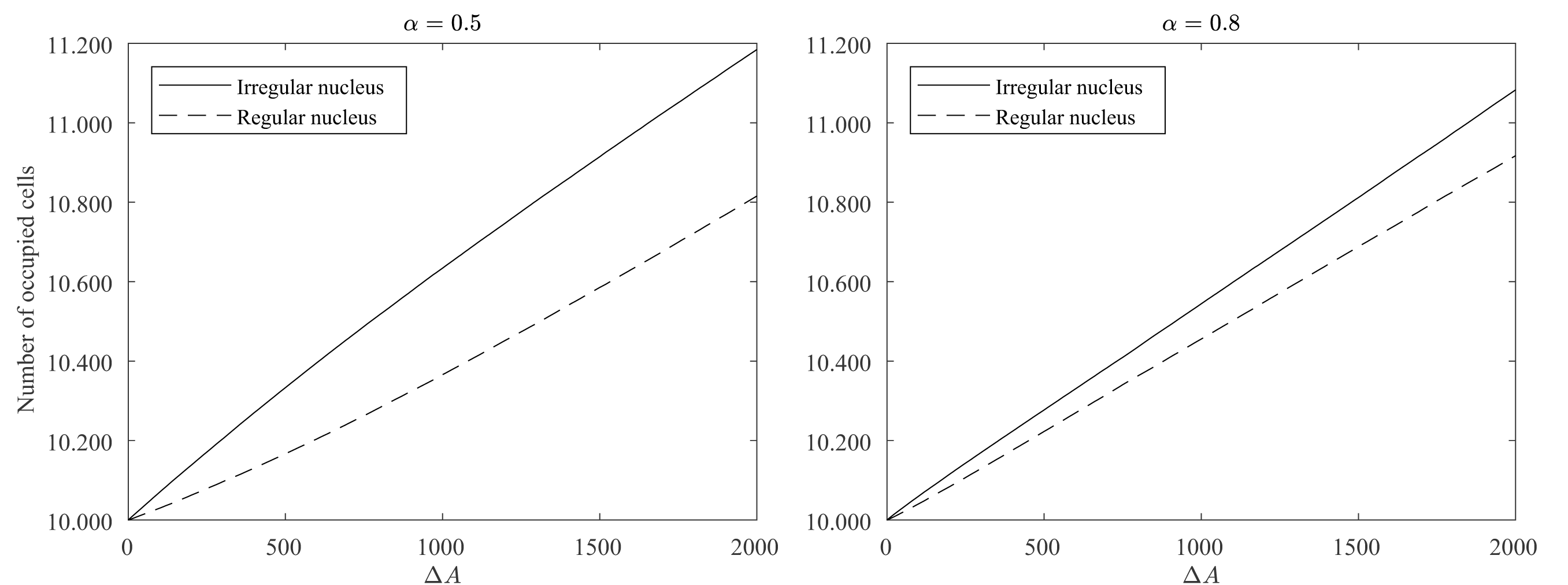}
  \caption{Evolution of the number of occupied cells for irregular and
    regular nuclei. We computed the mean of 100 simulations with the
    same parameter values as those used in the simulation of Figure
    \ref{fig:fractal_c05} with $\alpha =0.5$ (left) and $\alpha=0.8$
    (right).}
  \label{fig:fractal_ocupacion}
\end{figure}

\begin{figure}[htb]
  \centering
  \includegraphics[width = 0.6\linewidth]{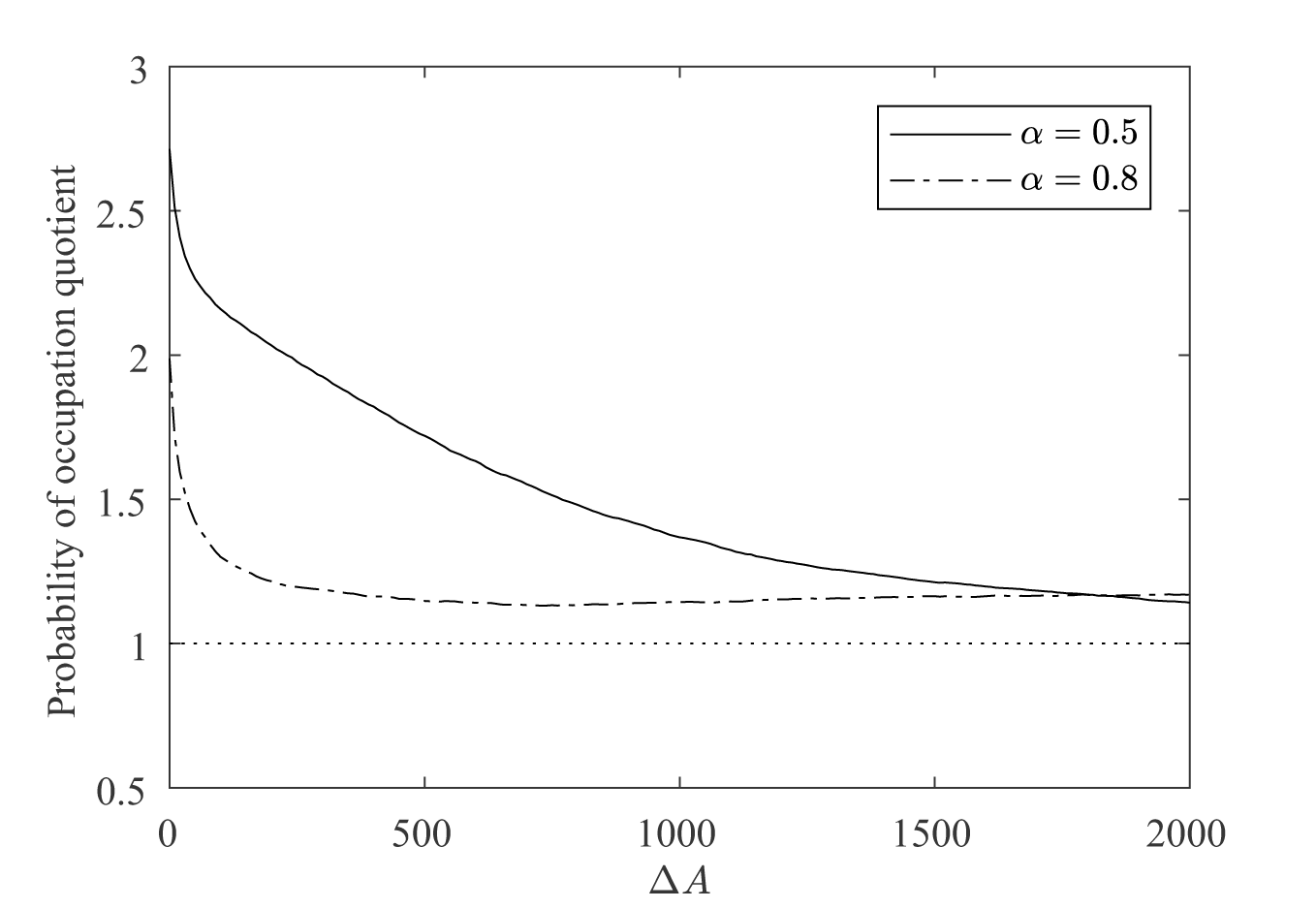}
  \caption{Evolution of the probability of occupation quotient, given
    by the probability of occupation of a cell adjacent to the
    irregular nucleus divided by the probability of occupation of a
    cell adjacent to the regular nucleus, with $\alpha =0.5$ and
    $\alpha=0.8$. 
}
  \label{fig:fractal_relacion_numboletos}
\end{figure}

For a quantitative estimation of the differences between the regular
and irregular nuclei, 100 simulations were conducted as described
above for $\alpha = 0.5$ and $\alpha = 0.8$. Figure
\ref{fig:fractal_ocupacion} shows how the growth of the regular
nucleus is slower than the irregular one. For example, in the case of
$\alpha = 0.5$, from the 2000 new cells grown by the simulation, only
around 800 were added to the regular nucleus, while the other 1200 were
added to the irregular one. It is also remarkable how, at the
beginning, the differences in the growth rates are higher, and as the
contour of the regular nucleus becomes more and more irregular, the
growth rates are equalized. This also becomes clear in Figure
\ref{fig:fractal_relacion_numboletos}, in which the ratio between the
probabilities of occupation of a cell adjacent to both nuclei is
plotted vs. the number of new deposited cells ($\Delta A$). It can be
seen how the curves are above 1, meaning that the irregular nucleus
has more occupation probability than the regular one and how the
curves decrease as the crystals grow, due to the fact that the
differences in regularity between the two nuclei are less marked. In
the long run, the quotients are asymptotically stabilized around a
value slightly greater than 1, because the irregular nucleus has grown
more in the initial stages of the simulation and these differences
remain over time. 

\bibliographystyle{unsrt}
\bibliography{references.bib}   

\end{document}